\documentstyle[12pt]{article}
\begin{document}
\newcommand{\cl}{\centerline}
\renewcommand{\theequation}{\arabic{equation}}
\newcommand{\beq}{\begin{equation}}
\newcommand{\eeq}{\end{equation}}
\newcommand{\bea}{\begin{eqnarray}}
\newcommand{\eea}{\end{eqnarray}}
\newcommand{\nn}{\nonumber}
\newcommand\pa{\partial}
\newcommand\un{\underline}
\newcommand\ti{\tilde}
\newcommand\pr{\prime}
\begin{titlepage}
\setlength{\textwidth}{5.0in} \setlength{\textheight}{7.5in}
\setlength{\parskip}{0.0in} \setlength{\baselineskip}{18.2pt}

\begin{flushright}
{\tt hep-th/0508022} \\
{\tt SOGANG-HEP315/05}
\end{flushright}

\vspace*{0.2cm}

\begin{center}
{\large\bf Improved BFT embedding having chain-structure}
\end{center}

\vskip 0.4cm

\begin{center}
{Yong-Wan {\sc Kim}$^{}$\footnote{\small Electronic address:
ywkim@sejong.ac.kr} and {\sc Ee} Chang-Young$^{}$\footnote{\small
Electronic address: cylee@sejong.ac.kr} \\
Department of Physics and Institute for Science and
Technology,\\
Sejong University, Seoul 143-747, Korea\\
\vskip 0.1cm
Seung-Kook {\sc Kim}$^{}$\footnote{\small Electronic
address:
skandjh@empal.com}\\
{Department of Physics, Seonam University, Chonbuk 590-170, Korea\\
\vskip 0.1cm
Young-Jai {\sc Park}$^{}$\footnote{\small Electronic
address:
yjpark@sogang.ac.kr}}\\
Department of Physics and Center for Quantum Spacetime, Sogang University, Seoul 121-742, Korea}\\
\end{center}

\vskip 0.4cm

\begin{center}
{\bf ABSTRACT}
\end{center}
\begin{quotation}
We newly revisit the gauge non-invariant chiral Schwinger model
with $a=1$ in view of the chain structure. As a result, we show
that the Dirac brackets can be easily read off from the exact
symplectic algebra of second-class constraints. Furthermore, by
using an improved BFT embedding preserving the chain structure, we
obtain the desired gauge invariant action including a new type of
Wess-Zumino term.

\vskip 0.4cm

\end{quotation}
\end{titlepage}

\newpage
\section{Introduction}
Gauge theories which play an important role in modern theoretical
physics belong to the class of the singular Lagrangian theories, and
there has been unceasingly interested in the quantization of these
constrained theories since Dirac's pioneering work \cite{Dirac}.
Batalin, Fradkin, and Vilkovsky \cite{bfv} had proposed a new kind
of quantization method for constraint systems, which is particularly
powerful for deriving a covariantly gauge-fixed action in
configuration space. Furthermore, Batalin, Fradkin, and Tyutin (BFT)
\cite{bft} had generalized this method introducing some auxiliary
fields in the extended phase space. After their works, several
authors \cite{KP,BRR,KBB,BanBanGh2} have systematically applied this
BFT method to various interesting models including the bosonized
Chiral Schwinger Model (CSM) \cite{csm}.

Recently, Shirzad and Monemzadeh \cite{ShirzadMone} have applied a
modified BFT method, which preserves the chain structure of
constraints \cite{Cabo-Chaichian}, to the bosonized CSM. However,
However, their constraint algebra is incorrect from the start.
Furthermore, they did not carry out further possible simplification
for the constraint algebra. As a result, although they have newly
applied the idea of the chain structure, they have not successfully
obtained a desired gauge invariant Lagrangian.

In this paper, we shall newly resolve this unsatisfactory situation
of the $a=1$ CSM by improving our BFT method \cite{KBB} and the
non-trivial application of the well-known technique of covariant
path integral evaluation \cite{bfv-app}. As results, we find the
gauge invariant quantum action revealing a new type of Wess-Zumino
(WZ) term \cite{KP}.

\section{Chain Structure of a Second-Class System}

First, let us consider a given canonical Hamiltonian $H_C$ with a
single primary constraint $\phi^{(1)}$ from the motivation of
analyzing the CSM to recapitulate the known chain-by-chain method
\cite{Cabo-Chaichian} briefly. The total Hamiltonian is defined as
\begin{eqnarray}
H_T=H_C+v\phi^{(1)},
\end{eqnarray}
where $v$ is a Lagrange multiplier to the primary constraint.
Then, from the consistency condition, {\it i.e.,}
$\dot\phi^{(1)}=0$ in the Hamiltonian formulation, we have
\begin{equation}
\label{pri-const} \dot\phi^{(1)}:=\{\phi^{(1)}, H_T\}=\{\phi^{(1)},
H_C\}+v\{\phi^{(1)}, \phi^{(1)}\}=0
\end{equation}
whose solutions are among three possible cases: (1) if
$\{\phi^{(1)},\phi^{(1)}\}\neq 0$, the Lagrange multiplier $v$ is
fixed, (2) if $\{\phi^{(1)}, H_C\}=0$ as well as
$\{\phi^{(1)},\phi^{(1)}\}=0$, the condition is identically
satisfied, and (3) if $\{\phi^{(1)}, H_C\} \neq 0$ as well as
$\{\phi^{(1)},\phi^{(1)}\}=0$, the condition generates a new
constraint as $\phi^{(2)} \equiv\{\phi^{(1)}, H_C\}$. Since we are
only interested in the last nontrivial case, we continue to
require the consistency condition on the constraint as
$\dot\phi^{(2)}=0$ with the condition of $\{\phi^{(2)}, H_C\} \neq
0$ as well as $\{\phi^{(2)},\phi^{(1)}\}=0$ as like in Eq.
(\ref{pri-const}), producing a next stage constraint $\phi^{(3)}
\equiv\{\phi^{(2)}, H_C\}$, and so forth. Let us assume that after
$(n-1)$-th step of the procedure, we obtain all constraints which
satisfy the following relations
\begin{eqnarray}
\label{primitive-const}
\{\phi^{(i)}, \phi^{(1)} \} &=& 0,  \nonumber\\
\{\phi^{(i)}, H_C\} &=& \phi^{(i+1)}, ~~~~~i=1,2,\cdot\cdot\cdot,
n-1.
\end{eqnarray}
While the consistency condition for the last constraint is explicitly written as
\begin{equation}
\dot\phi^{(n)}:=\{\phi^{(n)}, H_T\}=\{\phi^{(n)},
H_C\}+v\{\phi^{(n)}, \phi^{(1)}\}=0.
\end{equation}
Among three possible solutions for this consistency condition, let
us concentrate on the case that the Lagrange multiplier has been
finally determined as
\begin{equation}
v=-\frac{\{\phi^{(n)},H_C\}}{\{\phi^{(n)},\phi^{(1)}\}}
\end{equation}
with the requirement of $\{\phi^{(n)},\phi^{(1)}\}=\eta\neq 0$.
Note here that all the constraints $\phi^{(i)}$,
$(i=2,3,\cdot\cdot\cdot,n)$ are generated from the only one
primary constraint $\phi^{(1)}$ through the consistency
conditions, {\it i.e.}, the constraint set has a chain structure.

Furthermore, considering the Jacobi identities for the set of the
whole constraints and canonical Hamiltonian, $(\phi^{(i)}, H_C)$,
$i=1,2,\cdot\cdot\cdot,n)$, one can obtain
\begin{eqnarray}
\label{one-chain-const-rel-1}
\{\phi^{(i)},\phi^{(j)}\} &=& 0, ~~~~~~~~~~~~~i+j \leq n, \\
\label{one-chain-const-rel-2}
\{\phi^{(n-i+1)},\phi^{(i)}\}&=&(-1)^{i+1}\eta,~~~i=1,2,\cdot\cdot\cdot,n.
\end{eqnarray}
From the above relations, we have newly observed that if we define
the constraint algebra $\Delta_{\alpha\beta}\equiv
\{\phi_\alpha,\phi_\beta\}$ as usual, the relations
(\ref{one-chain-const-rel-1}) imply that all the matrix elements
above the off-diagonal component are simply zero. On the other hand,
the relations (\ref{one-chain-const-rel-2}) intimate that the
off-diagonal elements are not vanishing but alternating $\pm\eta$,
{\it i.e.,} $\Delta_{(1,n)}=-\eta$, $\Delta_{(2,n-1)}=+\eta$, etc,
implying that the system we are considering is of second-class (SC)
constraints. Moreover, one can also show that the total number of
the constraints are always even: that is to say, comparing two
relations both when $i=1$ and $i=n$, one obtain
$\{\phi^{(n)},\phi^{(1)}\}=\eta$ as well as
$\{\phi^{(1)},\phi^{(n)}\}=(-1)^{n+1}\eta$. Thus, the
antisymmetricity of the Poisson brackets gives rise to the even
number of constraints.

Our additional observation is in order: when $i+j\leq n$, among
the Jacobi identities
\begin{equation}
\{\phi^{(i)},\{\phi^{(j)}, H_C\}\}+\{\phi^{(j)},\{H_C,
\phi^{(i)}\}\}+ \{H_C, \{\phi^{(i)},\phi^{(j)}\}\} =0
\end{equation}
are reduced to
\begin{equation}
\{\phi^{(i)},\phi^{(j+1)}\}-\{\phi^{(j)}, \phi^{(i+1)}\}=0,
\end{equation}
where we have used the relations (\ref{one-chain-const-rel-1}) and
(\ref{one-chain-const-rel-2}). Therefore, as far as we consider
the upper part of the off-diagonal elements of the Dirac matrix
$\Delta_{\alpha\beta}$, the knowledge of the first low elements in
the matrix, generates all the remaining off-diagonal elements,
{\it i.e.,} downward left, starting from the very elements. That
is, when $i+j\leq n-1$, we have
\begin{eqnarray}
&& \{\phi^{(1)}, \phi^{(i+j)}\}=-\{\phi^{(2)}, \phi^{(i+j-1)}\} =
\cdot\cdot\cdot =\{\phi^{(j)}, \phi^{(i+1)}\}=
\nonumber \\
&& \cdot\cdot\cdot = \{\phi^{(i+j-1)}, \phi^{(2)}\}=
-\{\phi^{(i+j)}, \phi^{(1)}\} = 0.
\end{eqnarray}
Note that at this stage one can not still say anything on the
elements in the lower part, {\it i.e.,} below the off-diagonal
components in the Dirac's matrix $\Delta_{\alpha\beta}$, which is
the subject to discuss in below.

Now, let us consider all the constraints:
\begin{equation}
(\phi^{(1)}, \phi^{(2)},\cdot\cdot\cdot, \phi^{(k)}, \phi^{(k+1)},
\cdot\cdot\cdot, \phi^{(n-1)}, \phi^{(n)}),
\end{equation}
where $k=n/2$ and $n$ is even as proved before. To make the
discussion easier let us relabel the superscripts as
\begin{equation}
\phi^{*(k)}=(-1)^k \phi^{(n-k+1)}.
\end{equation}
As a result, we can construct the following $(k)$-pairs Poisson
brackets, which are all the off-diagonal elements of the Dirac
matrix $\Delta_{\alpha\beta}$ designated from the sum of the
superscripts to be $n+1$, as
\begin{eqnarray}
\{\phi^{(1)},\phi^{(n)}\}=-\eta,
~~~~~~~&\Rightarrow & ~~~~~~\{\phi^{(1)},\phi^{*(1)}\}=\eta,\nonumber\\
\{\phi^{(2)},\phi^{(n-1)}\}=\eta,
~~~~~~~&\Rightarrow & ~~~~~~~~\{\phi^{(2)},\phi^{*(2)}\}=\eta,\nonumber\\
&\cdot\cdot\cdot\cdot\cdot\cdot\cdot&\nonumber\\
\{\phi^{(k)},\phi^{(n-k+1)}\}=(-1)^k\eta, ~~~~~~&\Rightarrow &
~~~~\{\phi^{(k)},\phi^{*(k)}\}=\eta.
\end{eqnarray}

Now let us redefine the constraints as
\begin{equation}
\tilde{\phi}^{(1)} \equiv \phi^{(1)}, ~~~~ \tilde{\phi}^{*(1)}
\equiv \phi^{*(1)}.
\end{equation}
Then, through the Gram-Schmidt process, we can obtain new set of
the constraints
\begin{eqnarray}
\tilde{\phi}^{(k)} &=& \phi^{(k)} - \sum^{k-1}_{i=1}
\frac{\{\phi^{(k)},\tilde{\phi}^{*(i)}\}}{\{\tilde{\phi}^{(i)},
\tilde{\phi}^{*(i)}\}}\tilde{\phi}^{(i)},\nonumber\\
\tilde{\phi}^{*(k)} &=& \phi^{*(k)} - \sum^{k-1}_{i=1}
\frac{\{\phi^{*(k)},\tilde{\phi}^{*(i)}\}}{\{\tilde{\phi}^{(i)},
\tilde{\phi}^{*(i)}\}}\tilde{\phi}^{(i)},
\end{eqnarray}
which are satisfied with the following compact relations
\begin{eqnarray}
\{\tilde{\phi}^{(k)},\tilde{\phi}^{*(k')}\}&=&\delta^{kk'}\eta,\nonumber\\
\{\tilde{\phi}^{(i)},\tilde{\phi}^{(j)}\} &=& 0,~~~(i\neq j), \nonumber\\
\{\tilde{\phi}^{*(i)},\tilde{\phi}^{*(j)}\} &=& 0,~~~(i\neq j).
\end{eqnarray}
Therefore, we have shown that for any one-chain constrained system
we can obtain off-diagonalized Dirac matrix
$\Delta_{\alpha\beta}$, which is symplectic, in general. The
inverse of symplectic Dirac matrix is naturally used to define
usual Dirac brackets without much efforts. However, note here that
we should pay much attention to the Dirac matrix
$\Delta_{\alpha\beta}$ when there exist self-anticommuting matrix
components, {\it e.g.}, $\{\phi^{*(i)},\phi^{*(i)}\}\neq 0$,
$(i=1,2,\cdot\cdot\cdot,n/2-1)$. This arises in many cases
including the chiral boson, the chiral Schwinger, the
Maxwell-Chern-Simons, the self-dual models, and so on. In those
cases, the Dirac matrix usually contains a derivative term, which
is anticommuting itself, through a certain Poisson bracket in
constraint algebra. If then, one should first redefine some of
constraints giving a derivative term in constraint algebra to have
vanishing self-anticommuting matrix component in the Dirac matrix.

\section{CSM with Chain Structure}

As is well-known, the fermionic CSM with a regularization
ambiguity $a=1$ is equivalent to the following bosonized action
\cite{csm}
\begin{equation}
    S_{CSM}=\int d^2x \left[
                          -\frac{1}{4}F_{\mu \nu}F^{\mu \nu}
                     +\frac{1}{2}\partial_{\mu}\phi\partial^{\mu}\phi
                          +A_{\nu}(\eta^{\mu \nu}
                          -\epsilon^{\mu \nu})\partial_{\mu}\phi
              +\frac{1}{2} A_{\mu}A^{\mu} \right],
\end{equation}
where $\eta^{\mu\nu}=\mbox{diag(1,-1)}$, and $\epsilon^{01}=1$.
Comparing with the CSM with $a>1$, the number of constraints in
CSM with $a=1$ is four and the former has two. In order to
explicitly study the advantages of chain structure, it is very
instructive to apply the chain structure idea to the above
bosonized CSM with $a=1$.

The canonical momenta are given by
\begin{eqnarray}
    \pi^0&=& 0, \nonumber\\
    \pi^1&=& \dot{A}_1-\partial_{1}A_0 \equiv E,\nonumber \\
    \pi_\phi &=& \dot{\phi}+A_0-A_1 \equiv \pi,
\end{eqnarray}
where the overdot denotes the time derivative. The canonical
Hamiltonian is written as
\begin{eqnarray}
\label{can-H}
    H_C = \int dx \left[
            \frac{1}{2}E^2 + \frac{1}{2}\pi^2+ \frac{1}{2}(\partial_{1}\phi)^2
             + E\partial_1 A_0 + (\pi+ A_1+ \partial_1\phi)
             (A_1-A_0)\right].
\end{eqnarray}

Following Dirac's standard procedure \cite{Dirac}, one finds one
primary constraint
\begin{equation}
    \Phi_1 \equiv \pi^0 \approx 0,
\end{equation}
and three secondary constraints
\begin{eqnarray}
    \Phi_2 &\equiv& \partial_1 E + \pi+ \partial_1 \phi
              + A_1\approx 0 , \nonumber \\
    \Phi_3 &\equiv&  E \approx 0, \nonumber \\
    \Phi'_4 &\equiv& -\pi-\partial_1 \phi - 2 A_1+A_0 \approx 0.
\end{eqnarray}
These secondary constraints starting from the primary constraint
$\Phi_1$ are successively obtained by conserving the constraints
with respect to the total Hamiltonian in the usual Dirac scheme
\begin{equation}
    H_T = H_C + \int dx~ v\Phi_1,
\end{equation}
where $v$ denotes a Lagrange multiplier, which is fixed as follows
\begin{equation}
        v =\partial_1 \pi +\partial^2_1 \phi + 2E+2\partial_1 A_1.
\end{equation}
No further constraints are generated in this procedure showing that
the CSM with $a=1$ belongs to one-chain, and all the constraints are
fully second-class (SC) constraints.

It is appropriate to comment that in the work of Shirzad at al.
there should be a term in their constraint algebra
\cite{ShirzadMone} in which $\Delta_{44}$ is not simply zero but
$2\partial_x\delta(x-y)$. This term is in itself antisymmetric,
which case is excluded from the general consideration of chain
structure as seen in Sec. [2]. Therefore, before we proceed to
further, we have to redefine this constraint by making use of the
other constraints.

Now, let us redefine $\Phi'_4$ as $\Phi_4$ by using $\Phi_1$ as
follows
\begin{eqnarray}
\label{mod-const4}
        \Phi_4 &\equiv&
                \Phi'_4 + \partial_1 \Phi_1 \nonumber\\
         &=& - \pi - \partial_1\phi -2A_1 + A_0+\partial_1 \pi^0 .
\end{eqnarray}
Note that the redefined constraint is still SC one. However, the new
set of the constraints $\Phi_i (i=1,\cdot\cdot\cdot,4)$ satisfies
more simplified form of the SC constraints algebra
\begin{eqnarray}
\label{const-alg}
    \Delta_{ij}(x,y)
                &\equiv&  \{ \Phi_i(x), \Phi_j(y) \}
                                            \nonumber\\
                &=&  \left( \begin{array}{cccc}
                                          0   &  0   &  0    &  -1
\\
                                          0   &  0   &  1    &  0
\\
                                          0   & -1   &  0    & 2     \\
                                          1   &  0   & -2    &  0
                                 \end{array} \right)
  \delta(x -y),
\end{eqnarray}
which is a constant and antisymmetric matrix. Moreover, all the
matrix elements above the off-diagonal components are simply zero,
and the off-diagonal elements are alternating $\mp 1$.

Furthermore, by following the previous discussion on the chain
structure, we can make use of the Gram-Schmidt process to define
all the constraints as orthogonalized ones
\begin{equation}
\label{final-const}
\Omega_i=(\tilde{\phi}^{(1)},\tilde{\phi}^{(2)},
\tilde{\phi}^{*(1)},\tilde{\phi}^{*(2)}),
\end{equation}
where
\begin{eqnarray}
\label{new-const}
\tilde{\phi}^{(1)}&=&\Phi_1,~~~ \tilde{\phi}^{*(1)} = - \Phi_4,\nonumber\\
\tilde{\phi}^{(2)}&=&\Phi_2,~~~\tilde{\phi}^{*(2)}=\Phi_3+2\Phi_1.
\end{eqnarray}
Then, the final form of the Dirac matrix turns out to be
symplectic matrix as
\begin{eqnarray}
\label{symp-algebra} \widetilde{\Delta}_{ij}(x,y) &\equiv&
\{\Omega_i(x), \Omega_j(y)\} \nonumber\\
&=&\left(\begin{array}{cc}
         O & I \\
         -I & O
        \end{array} \right)\delta(x-y)
        \equiv J\delta(x-y),
\end{eqnarray}
where $I$ represents $(2\times 2)$ identity matrix.

Therefore, we have completely converted the Dirac matrix of the
CSM into the desired symplectic one by analyzing the chain
structure of the model systematically, which structure is of
one-chain. Now, as we know, the Dirac brackets can be obtained
from the symplectic matrix (\ref{symp-algebra}) without much
efforts.

It seems appropriate to comment on the previous work
\cite{ShirzadMone} of Shirzad et al. They used the constraint
$\Phi'_4$ wrongly to obtain the constraint algebra
(\ref{const-alg}) which constraint gives a derivative term in
$\Delta_{44}$ as stated above. In other words, the constraint
$\Phi'_4$ gives a self-anticommuting matrix element in the
constraint algebra, and in order to eliminate the corresponding
self-anticommuting element for further analysis they should have
modified the constraint $\Phi'_4$ to $\Phi_4$ as shown in
Eq.(\ref{mod-const4}). Then, one can obtain the correct constraint
algebra (\ref{const-alg}).

\section{Improved BFT Embedding}

At least for the one-chain system we have shown that the Dirac
matrix can be finally reduced to the desired symplectic form using
the fact that we can make the SC constraints to be paired with and
apply the Gram-Schmidt process to make them off-diagonal. As a
result, we have shown that using the chain structure of constrained
system has a great advantage of finding the Dirac brackets of SC
system easier than finding them based on the standard Dirac scheme.
On the other hand, the BFT algorithm is known to systematically
convert gauge non-invariant theory into gauge invariant one by
introducing auxiliary fields, and with the improved BFT formalism
preserving the chain structure we will later quantize the bosonized
CSM in the path integral framework.

The basic idea on the usual BFT algorithm is to convert the SC
constraints $\Omega_i$ into fully first-class (FC) ones $\tau_i$ by
introducing auxiliary fields $\eta^i$ as
\begin{eqnarray}
\label{first-class} \tau_i = \Omega_i + X_{ij}\eta^j +{\cal O}
(\eta^{(2)}),
\end{eqnarray}
where $\Omega_i$ is given by Eq.(\ref{final-const}) and requiring
that fully FC constraints satisfy the strongly involutive Poisson
brackets relations
\begin{equation}
\label{inv-rel} \{\tau_i, \tau_j\}=0.
\end{equation}
The strongly involutive Poisson brackets relation (\ref{inv-rel})
gives the following set of equations  on the zero-th order of
$\eta$ to be solved for $X_{ij}$ and $\omega^{ij}$ as
\begin{equation}
\label{couped-eq} \widetilde{\Delta}_{ij}(x,y)+\int dz_1 dz_2
X_{ik}(x,z_1)\omega^{kl}(z_1,z_2)X_{jl}(z_2,y)=0,
\end{equation}
where $\omega^{ij}$ is given by $\{\eta^i, \eta^j\}=\omega^{ij}$ for
the newly introduced auxiliary fields. Therefore, by choosing a set
of proper Ansatz for $\omega^{ij}$, $X_{ij}$, we can solve the set
of equations (\ref{couped-eq}) which means that we can successfully
convert the SC constraints into fully FC ones (\ref{first-class}).

However, the practical finding of FC quantities including the
Hamiltonian is not so straightforward in real world. The
difficulties arise from the fact that there is no guided proper way
to choose the matrix elements for $X_{ij}$ and $\omega^{ij}$ in Eq.
(\ref{couped-eq}). One has learned by experience that for any
different choices of $X_{ij}$ and $\omega^{ij}$ the corresponding FC
functions including the constraints and Hamiltonian are equivalent
to each other up to the existing constraints. In this respect, it
would have great advantage of choosing $X_{ij}$ and $\omega^{ij}$ as
simple as possible for further analysis. In our case, since we have
obtained the simplest expression for the Dirac matrix
$\widetilde{\Delta}_{ij}(x,y)$ in its symplectic form by using the
chain structure of constrained system, the natural candidates for
the solutions of $X_{ij}$ and $\omega^{ij}$ shall be simply given by
$X_{ij}=\widetilde{\Delta}_{ij}$ and
$\omega^{ij}=-\widetilde{\Delta}_{ij}$ where
$\widetilde{\Delta}_{ij}$ is now the symplectic matrix
(\ref{symp-algebra}). Moreover, the simple choices of
$X_{ij}=\widetilde{\Delta}_{ij}$ and
$\omega^{ij}=-\widetilde{\Delta}_{ij}$, {\it i.e.}, the exactly
symplectic matrices, may further simplify the explicit construction
of FC quantities since according the usual BFT formalism they would
be proportional only to the second powers in the auxiliary fields.
This would help to easily construct FC functions for complicated
theories including non-abelian models not in infinite but in finite
powers in the auxiliary fields.

Now let us come back to the CSM with $a=1$ case, and start from
defining the solutions of $\omega^{ij}=-\widetilde{\Delta}_{ij}$
and $X_{ij}=\widetilde{\Delta}_{ij}$ in Eq. (\ref{couped-eq}) as
\begin{eqnarray}
\omega^{ij}(x,y)&=&-J\delta(x-y),\nonumber\\
X_{ij}(x,y)&=&J\delta(x-y).
\end{eqnarray}
Then, the fully FC constraints are explicitly written by
\begin{eqnarray}
\tau_1&=&\Omega_1+\eta^3=\pi^0+\eta^3,\nonumber \\
\tau_2&=&\Omega_2+\eta^4=\partial_1E+\partial_1\phi+\pi+A_1+\eta^4,\nonumber\\
\tau_3&=&\Omega_3-\eta^1=\pi+\partial_1\phi+2A_1-A_0-\eta^1,\nonumber\\
\tau_4&=&\Omega_4-\eta^2=E+2\pi_0-\eta^2.
\end{eqnarray}
Note that the higher order correction terms ${\eta}^{(n)} ~(n \geq
2)$ in Eq. (30) automatically vanish as a consequence of the proper
choice (33). Therefore, we have all the FC constraints with only
$\eta^{(1)}$ contributing in the series (30) in the extended phase
space.

On the other hand, the construction of FC quantities can be done
along similar lines as in the case of the constraints, by
representing them as a power series in the auxiliary fields and
requiring
\begin{equation}
\label{first-construct} \{\tau_i, \widetilde{F}\}=0
\end{equation}
subject to the condition $\widetilde{F}({\cal O}; \eta^i = 0) =
F$. Here $F({\rm or,~}{\cal O})$ is a quantity (or, ~variables) in
the original phase space, while $\widetilde{F}$ is a quantity in
the extended phase space.

Then, we obtain the proper BFT physical variables as
\begin{eqnarray}
\label{phy-fields} \widetilde{A}^{\mu} &=& A^{\mu} + A^{\mu (1)}
                     = (A^{0} +{\eta}^1 +\partial_{1} \eta^{3} +2\eta^{4} ,
      A^{1} - \partial_{1} \eta^{2}  + \eta^{4} )  \nonumber \\
\widetilde{\pi}^{\mu} &=& \pi^{\mu} + \pi^{\mu (1)}
                      = (\pi^{0}+ \eta^{3},
                         \pi^{1}- \eta^{2}-2\eta^{3} ), \nonumber \\
\widetilde{\phi} &=& \phi + \phi^{(1)}
                  = \phi + {\eta}^2 + \eta^{3}, \nonumber \\
\widetilde{\pi}_\phi &=& \pi_{\phi} + \pi_\phi^{(1)}
                    = \pi_{\phi} + \partial_{1} \eta^{2}+ \partial_{1} \eta^{3}.
\end{eqnarray}
Meanwhile, the non-vanishing Poisson brackets of the physical
fields (\ref{phy-fields}) in the extended phase space are directly
read off as
\begin{eqnarray}
&& \{\tilde{A}_0(x), \tilde{A}_0(y)\}=2\partial^x_1\delta(x-y),~
\{\tilde{A}_0(x), \tilde{\phi}(y)\}=\delta(x-y),\nonumber\\
&& \{\tilde{A}_0(x), \tilde{\pi}(y)\}=-\partial^x_1\delta(x-y),~~
\{\tilde{A}_1(x),\tilde{A}_1(y)\}=2\partial^x_1\delta(x-y),\nonumber\\
&& \{\tilde{A}_1(x), \tilde{\phi}(y)\}=\delta(x-y),~~~~~~~~
\{\tilde{A}_1(x), \tilde{\pi}(y)\}=-\delta(x-y),\nonumber\\
&& \{\tilde{\phi}(x), \tilde{\pi}(y)\}=\delta(x-y).
\end{eqnarray}
These are exactly the same Dirac brackets for consistent
quantization in the traditional Dirac scheme where the quantum
brackets are usually found through a tedious algebraic
manipulation.

The FC Hamiltonian can be also obtained either by following the
prescription (\ref{first-construct}) directly or by noting that any
functional of FC fields such as $\tilde{{\cal
F}}=(\tilde{A}^\mu,\tilde{\pi}^\mu,\tilde{\phi},\tilde{\pi}_\phi$)
will be FC. As results, we have obtained the FC Hamiltonian from the
former method as
\begin{eqnarray}
\label{fisrt-H} \widetilde{H} = H_C+H^{(1)}+H^{(2)},
\end{eqnarray}
where
\begin{eqnarray}
H^{(1)}&=& \int dx
\left[-\eta^1(\partial_1E+\partial_1\phi+\pi+A_1)-\eta^2E\right.\nonumber\\
&&~~~~~~~\left.-\eta^3(2E+\partial_1A_1-\partial^2_1E)-\eta^4
(\pi+\partial_1\phi+2\partial_1E+A_0)
\right],\nonumber\\
H^{(2)}&=&\int
dx\left[-\eta^1\eta^4+\frac{1}{2}(\eta^2)^2+2\eta^2\eta^3-
\frac{1}{2}\eta^3\partial_1\eta^4
+\frac{1}{2}\eta^4\partial_1\eta^3\right.\nonumber\\
&&~~~~~~~\left.+2(\eta^3)^2-\eta^3\partial^2_1\eta^3-(\eta^4)^2\right],
\end{eqnarray}
where the superscripts denotes the power of the auxiliary fields
$\eta^i$ and we note that higher power terms greater than
$\eta^{(2)}$ in the auxiliary fields do not appear in the FC
Hamiltonian. One can easily show that the FC Hamiltonian
(\ref{fisrt-H}) is exactly equivalent to the one obtained from the
replacement of the original fields in the canonical Hamiltonian
(\ref{can-H}) by the physical BFT fields.

Finally, one can easily write down the desired Hamiltonian
$\widetilde{H}_T$, which exactly preserves the chain structure of
the constraint system in the extended phase space,
\begin{equation}
\label{tot-fin-ham} \widetilde{H}_T = \widetilde{H}+\sum^{3}_{i=1}
\eta^i \widetilde{\Omega}_{i+1}
\end{equation}
equivalent to the strongly involutive Hamiltonian $\widetilde{H}$.

Since in the Hamiltonian formalism the FC constraint system
indicates the presence of a local symmetry, this completes the
operatorial conversion of the original SC system with Hamiltonian
$H_C$ and constraints $\Phi_i$ into the FC ones in the extended
phase space with Hamiltonian $\widetilde H$ and constraints
$\Omega_i$ by using the property (\ref{first-construct}).

\section{Gauge Invariant Quantum Lagrangian}

In this section, we consider the partition function of the model
in order to extract out the gauge invariant quantum Lagrangian
corresponding to $\widetilde{H}_T$. Our starting partition
function is given by the Faddeev-Popov (FP) formula \cite{FP} as
follows
\begin{equation}
    Z =  \int  {\cal D} A_\mu  {\cal D} \pi^\mu
               {\cal D} \phi {\cal D} \pi
                 \prod_{i,j = 1}^{4}{\cal D} \eta^i
                 \delta(\tilde{\Omega}_i )\delta(\Gamma_j )
                 \det \mid \{\tilde{\Omega}_i,\Gamma_j \} \mid
                 e^{iS},
\end{equation}
where
\begin{equation}
        S  =  \int d^2x \left(
                \pi^\mu {\dot A}_\mu
                + \pi {\dot \phi}
                + \frac{1}{2}\sum^{4}_{i=1}\eta^i\omega_{ij}\dot{\eta}^j
                 - \widetilde{\cal H}_T
                   \right)
\end{equation}
with the Hamiltonian density $\widetilde{\cal H}_T$. The gauge
fixing conditions $\Gamma_i$ should be chosen so that the
determinant occurring in the functional measure is non-vanishing.
Moreover, $\Gamma_i$ may be assumed to be independent of the
momenta so that these are considered as the FP type gauge
conditions \cite{KP,BRR,KBB}.

Following the standard path integral method, one can obtain
the following gauge invariant action $S_F$
\begin{eqnarray}
    S_{F}       &=& S_{CSM} + S_{WZ} + S_{NWZ} ~;      \nonumber \\
    S_{CSM} &=&  \int d^2x~ \left[
                -\frac{1}{4}F_{\mu \nu}F^{\mu \nu}
                +\frac{1}{2}\partial_{\mu}\phi\partial^{\mu}\phi
                + A_{\nu}(\eta^{\mu \nu}
                -\epsilon^{\mu \nu})\partial_{\mu}\phi
        +\frac{1}{2}A_{\mu}A^{\mu}~\right], \nonumber \\
    S_{WZ} &=&  - \int d^2x  ~\frac{1}{2}\eta^3\epsilon_{\mu\nu}F^{\mu\nu}, \nonumber \\
    S_{NWZ} &=&  \int d^2x
            \left[
             A_0(\dot{\eta}^3-\partial_1\eta^3+2\eta^4)-A_1(\partial_1\eta^3
            +\eta^4)+(\dot{\phi}+\partial_1\phi)(\dot{\eta}^3+\eta^4)
           \right.
            \nonumber\\
            &&\left.+\partial_1\eta^3(\dot{\eta}^3-\partial_1\eta^3-\eta^4)
            +\frac{1}{2}(\dot{\eta}^3+\eta^4)^2+(\eta^4)^2
            -\frac{1}{2}(\dot{\eta}^4-\partial^2_1\eta^3-2\partial_1\eta^4)^2
             \right.\nonumber\\
\end{eqnarray}
The desired final action $S_F$ is invariant under the gauge
transformations as
\begin{eqnarray}
\delta A_\mu = \partial_\mu \Lambda,~\delta \phi = \Lambda,
~\delta \eta^3 = - \Lambda,~\delta \eta^4=0,
\end{eqnarray}
where the new type of WZ action $S_{NWZ}$ in itself is invariant
under the above transformations, which means that this term is not
related to the gauge symmetry. The resultant action has not only the
well-known WZ term $S_{WZ}$ canceling the gauge anomaly, but also a
new type of WZ term $S_{NWZ}$, which is irrelevant to the gauge
symmetry but is needed to make the SC system into the fully FC one
analogous to the case of the CS model \cite{KP}.

\section{Conclusion}

In conclusion, we have revisited the chain structure
\cite{ShirzadMone} analysis for the nontrivial $a=1$ bosonized
CSM, which belongs to one-chain system with one primary and three
secondary constraints. In this chain structure formalism, we have
newly defined the second-class constraints as the proper
orthogonalized ones, and then have successfully converted the
Dirac matrix into the symplectic one. As a result, we have
resolved the unsatisfactory situation of the $a=1$ CSM in the
incomplete previous work \cite{ShirzadMone}. Furthermore, based on
our improved BFT method preserving the chain structure in the
extended phase space, we have found the desired gauge invariant
quantum action.

Through further investigation, it will be interesting to apply
this newly improved BFT method to non-Abelian cases
\cite{self-dual} as well as an Abelian four-dimensional anomalous
chiral gauge theory \cite{Raja}, which seem to be very difficult
to analyze within the framework of the original BFT formalism
\cite{BRR}.

\section*{Acknowledgments}

The work of Y.-W. {\sc Kim} was supported by the Korea Research
Foundation, Grant No. KRF-2002-075-C00007. The work of {\sc Ee}
C.-Y. was supported by KOSEF, Grant No. R01-2000-000-00022-0. The
work of Y.-J. {\sc Park} was supported by Center for Quantum
Spacetime through KOSEF, Grant No. R11-2005-021.

\end{document}